\documentclass[10pt,a4paper]{article}

\usepackage{amssymb,amsmath,mathtools} % Math
\usepackage{graphicx} % Graphics
\usepackage{bbm} % Double lined letters
\usepackage{tikz} % TikZ ist kein Zeichenprogramm, oder?

\usepackage[font=small,labelfont=bf]{caption} % Captions

\usepackage{enumerate} % Lists

\usepackage[hmargin=0.12\paperwidth,vmargin=0.2\paperwidth,bindingoffset=0cm]{geometry}

\usepackage{titlesec}

\titleformat*{\section}{\large\bfseries}
\titleformat*{\subsection}{\bfseries}

\newcommand{\one}{\mathbf{I}} % One?

\renewcommand{\Re}{\operatorname{Re}} % Real part
\newcommand{\p}{\partial} % Partial

\newcommand{\E}{\mathbb E}

\renewcommand{\phi}{\varphi} % I prefer varphi
\renewcommand{\epsilon}{\varepsilon} % I prefer varepsilon

\DeclareMathOperator{\tr}{Tr} % Trace
\DeclareMathOperator*{\diag}{diag} % Diagonal
\DeclareMathOperator{\gGl}{GL} % General linear group
\DeclareMathOperator{\gO}{O} % Orthogonal group
\DeclareMathOperator{\gU}{U} % Unitary group

\DeclarePairedDelimiter{\abs}{\lvert}{\rvert} % Absolute value
\DeclarePairedDelimiter{\norm}{\lVert}{\rVert} % Norm
\DeclarePairedDelimiter{\normOrd}{\vcentcolon}{\vcentcolon} % Normal ordering

\usepackage{hyperref}   % Internal hyperlinks
\hypersetup{
pdfborder={0 0 0},      % No borders around internal hyperlinks
}

\title{\bfseries\Large Isotropic Brownian motions over complex fields\\ as a solvable model for May--Wigner stability analysis}
\author{J.~R.~Ipsen${}^1$ and H. Schomerus${}^2$}
\date{\today}

\usepackage{lipsum}

\begin{document}	

\maketitle

\begin{center}
 ${}^1$ARC Centre of Excellence for Mathematical and Statistical Frontiers,\\ School of Mathematics and Statistics, The University of Melbourne, VIC 3010, Australia\\[\baselineskip]
 ${}^2$Department of Physics, Lancaster University, LA1 4YB Lancaster, United Kingdom
\end{center}

\begin{abstract}
\noindent
We consider matrix-valued stochastic processes known as isotropic Brownian motions, and show that these can be solved exactly over complex fields.
While these processes appear in a variety of questions in mathematical physics, our main motivation is their relation to a May--Wigner-like stability analysis, for which we obtain a stability phase diagram.
The exact results establish the  full joint probability distribution of the finite-time Lyapunov exponents, and may be used as a starting point for a more detailed analysis of the stability-instability phase transition.
Our derivations rest on an explicit formulation of a Fokker--Planck equation for the Lyapunov exponents. This formulation happens to coincide with an exactly solvable class of models of the Calgero--Sutherland type, originally encountered for a model of phase-coherent transport. The exact solution over complex fields describes a determinantal point process of biorthogonal type similar to recent results for products of random matrices, and is also closely related to Hermitian matrix models with an external source.
\end{abstract}

\section{Introduction}
\label{sec:intro}

A close link between stochastic calculus and random matrix theory exists since the seminal work of Dyson~\cite{Dyson:1962brown}, who introduced a stochastic process on matrices which greatly simplifies the task to
implement various canonical ensembles.
Since its introduction, this method, now known as Dyson Brownian motion, has found a wide variety of applications in both physics and mathematics, see e.g.~\cite{Katori:2016}.
Two important examples of matrix-valued stochastic models
in physics concern the passive advection in fully developed turbulence~\cite{Kraichnan:1974,MY:1975,LeJan:1985,BH:1986,CFK:1994,GK:1996,FGV:2001}, where one is interested in the stability of the flow, and the phase-coherent transport through disordered wires~\cite{MPK:1988,MS:1991,MC:1992,BR:1994,Frahm:1995,Beenakker:1997,MBF:1999,TBFM:2001}, where one encounters Anderson localization. While Dyson's model has additive noise, these models have multiplicative noise, which gives rise to scaling regimes with statistics that differ from the usual Wigner--Dyson type. Not surprisingly, there are numerous close connections between these two classes of models with multiplicative noise \cite{LGP:1988,CPV:1993}.

In this work, we identify  a similarly close relation starting from a paradigmatic model, sometimes referred to as isotropic Brownian motions~\cite{LeJan:1985}, isotropic stochastic flows~\cite{BH:1986}, or matrix-valued multiplicative diffusions~\cite{GJJN:2003}. For motivation, we consider this model in the context of a May--Wigner-like stability analysis of large complex systems, as they occur in a wide variety of settings in physics, biology and beyond.
We show that the finite-time Lyapunov exponents share the same statistics as those appearing in the transport through a quantum wire with chiral symmetry. This connection is surprising inasmuch as the exponents for the quantum wire are  constrained by a condition related to flux conservation, which does not have a counterpart for the isotropic Brownian motion. The relation between the models allows us to find an exact solution for motions over complex fields, owing to the fact that the statistics in the corresponding transport problem is determined by a Hamiltonian of the  Calgero-Sutherland type. Furthermore, we identify connections to products of random matrices, including  Newman's square law~\cite{Newman:1986} and biorthogonal ensembles~\cite{AI:2015,Ipsen:2015thesis}, as well as Hermitian matrix models with an external source~\cite{BH:1996,ZinnJustin:1997,ABK:2005}. These findings help us to establish a simple stability phase diagram, expressed in terms of an effective parameter that incorporates the rotational and gradient components of the flow, and defines a convenient starting point for more detailed investigations of the phase transition.

First, let us recall some basics about May--Wigner-like stability analysis. We consider a generic autonomous dynamical system given by
\begin{equation}\label{intro:ODE}
\frac{dy_i(t)}{dt}=f_i(y_1(t),\ldots,y_N(t)),\qquad i=1,\ldots,N,
\end{equation}
where $\{f_i\}$ is some family of smooth functions and the dimension $N$ is assumed to be large. Introducing a small perturbation $y(t)\mapsto y(t)+\epsilon\tilde y(t)$ leads to the linearised problem
\begin{equation}\label{intro:linear}
\frac{d\tilde y_i(t)}{dt}=\sum_{j=1}^N\tilde y_j(t)\frac{\p f_i}{\p y_j}+O(\epsilon).
\end{equation}
If the expansion is about a fixed point the gradient matrix $[\p f_i/\p y_j]$ is independent of time. It is then canonical to write
${\p f_i}/{\p y_j}=-\mu\,\delta_{ij}+\sigma J_{ij}$,
where $J_{ij}$ incorporates the coupling between $\tilde y_i$ and $\tilde y_j$, while $\sigma$ sets an overall coupling strength and $\mu$ sets a drift which for positive values may be interpreted as the relaxation rate at $\sigma=0$.

For sufficiently complex systems it is reasonable to assume  (at least as a toy model) that $J=[J_{ij}]$ is a ``maximally random'' matrix, up to symmetries stemming from physical considerations. Such models were introduced in a seminal paper by May~\cite{May:1972}. The original paper was mainly focussing on the stability of large ecosystems, where it challenged the (at the time) common folklore that the stability of complex systems increases with the size of the system. Similar ideas as those introduced in~\cite{May:1972} have since been applied to many other systems, including e.g. machine learning~\cite{GF:2013}, finance~\cite{HM:2011}, and neural networks~\cite{SCS:1988}. There is also a close connection to what is known as random landscapes~\cite{Fyodorov:2004,Fyodorov:2013,FK:2015}.

Rather than an expansion around a fixed point, we imagine to evaluate the gradient matrix along a trajectory, see e.g.~\cite{CPV:1993}. In this case, the gradient matrix  depends on the position on the trajectory and therefore indirectly on time. With May's model in mind, we write
\begin{equation}\label{intro:SDE}
 \frac{d\tilde y_i(t)}{dt}=\sum_{j=1}^N(-\mu\,\delta_{ij}+\sigma\,J_{ij}(t))\tilde y_j(t)
\end{equation}
with some time-dependent coupling matrix $J_{ij}(t)$, to be further specified below. 
This is such that we have an attractive trajectory with relaxation rate $\mu$ when the coupling strength vanishes, i.e. when $\sigma=0$. Our main assumption is that the time-dependent coupling $J_{ij}(t)$ can be treated as noise, which we let be white in time. Such an approximation is valid in the limit of strong chaoticity, where the correlation time is short.

We therefore consider the system~\eqref{intro:SDE} driven by Gaussian white noise, with the only constraint that the system must be \emph{isotropic}. Originally, this type of process arose in models for passive advection in fully developed turbulence (here isotropy appears as an axiom stemming from K41 theory), see e.g.~\cite{MY:1975,FGV:2001}. The insight that temporal correlations may be neglected for turbulent velocity fields in the limit of high Reynolds numbers dates back to Kraichnan~\cite{Kraichnan:1968}. This type of isotropic models has also been considered by the mathematical community, see e.g.~\cite{LeJan:1985,Newman:1986,BH:1986,NRW:1986}. Later methods originating from random matrix theory and quantum field theory have been applied as well~\cite{GJJN:2003,GK:1996}.

We recall that the white-noise limit of processes like~\eqref{intro:SDE} depends on the regularisation  (the one-dimensional case, which is a geometric Brownian motion, provides a well-known example). Thus one needs to choose an interpretation, or alternatively impose additional constraints (see e.g. the appendix in~\cite{FGV:2001}).
For consistency with the existing literature~\cite{LeJan:1985,Newman:1986} we choose the Stratonovich interpretation as our starting point,
\begin{equation}\label{intro:stratonovich}
d\tilde y_i(t)=-\mu\,\tilde y_i(t)dt+\sigma\sum_{j=1}^N\tilde y_j(t)\circ dB_{ij}(t),\qquad i=1,\ldots,N,
\end{equation}
where $B(t)=[B_{ij}(t)]$ is a matrix-valued Brownian motion assumed to be statistically invariant under unitary similarity transformations. This is the type of process that we refer to as \emph{isotropic Brownian motion}.

For the practical implementation% and connection to the exactly solvable transport problem
, on the other hand, it will be convenient to work in It\^o convention, in  which the process~\eqref{intro:stratonovich} becomes
\begin{equation}\label{intro:ito}
d\tilde y_i(t)=-\mu\,\tilde y_i(t)dt+\sigma\sum_{j=1}^N\tilde y_j(t) dB_{ij}(t)
+\frac12\sigma^2\sum_{j,k=1}^N\tilde y_j(t)\E[dB_{ik}(t)dB_{kj}(t)].
\end{equation}
We recall that formally $\E[dB_{ik}(t)dB_{kj}(t)]\propto dt$. We note that the difference between Stratonovich and another interpretation (such as It\^o) is merely an additional spurious drift term which, in principle, could be absorbed into the constant $\mu$. 

Before we go on to the main results of this paper, let us recall a few basic properties of  isotropic Brownian motions.
Given an initial condition $y(0)$, the solution to the system~\eqref{intro:ito} may in matrix notation be written as
\begin{equation}\label{intro:solution}
y(t)=\Pi_\mu(t)y(0),
\end{equation}
where $\Pi_\mu(t)$ is a random evolution matrix, which itself satisfies an It\^o equation similar to~\eqref{intro:ito},
\begin{equation}
d\Pi_\mu(t)=\big(-\mu\,dt+\sigma dB(t)+\tfrac12\sigma^2\E[dB(t)dB(t)]\big)\Pi_\mu(t),
\end{equation}
with initial condition $\Pi_\mu(0)=\one$. Moreover, with $\Pi(t,s)$ denoting the evolution from time $s$ to time $t$, we have the almost-sure property $\Pi_\mu(t,s)\Pi_\mu(s,0)=\Pi_\mu(t,0)$ for all $t\geq s\geq0$. Per definition, the solution is given as an ordered product,
\begin{equation}\label{intro:matrix-product}
\Pi_\mu(t)=\lim_{n\to\infty}
\normOrd{\prod_{k=1}^n\big(1+(\mu+\tfrac12\sigma^2\E[(X^{(k)})^2])\delta t+\sigma\,X^{(k)}\delta t^{1/2}\big)},
\end{equation}
where $\delta t:=t/n$ is the time regularisation and $\{X^{(k)}\}$ is a family of independent random matrices given as the Brownian increments, $X^{(k)}\delta t^{1/2}=B(k\delta t)-B((k-1)\delta t)$. We could, of course, also write the random evolution matrix as a time-ordered exponential, but we will not need that formulation here. It follows that finding the random evolution matrix~\eqref{intro:ito} boils down to evaluating a product of independent random matrices; a topic which in the discrete setting has received considerable recent attention~\cite{AKW:2013,AIK:2013,Forrester:2013,Kargin:2014,ABK:2014,Ipsen:2015,Forrester:2015}.

In this paper, we consider evolutions over real as well as complex vector fields; thus, the random matrix $X(=X^{(k)})$ may be either real or complex. Adopting notation from random matrix theory, the real and complex case will be denoted by an index $\beta=1$ or $\beta=2$, respectively, whenever the distinction is important.

Isotropy implies that $U^{-1}XU\stackrel{d}{=}X$ for all rotations $U$, i.e. $U\in\gO(N)$ for $\beta=1$ and $U\in\gU(N)$ for $\beta=2$. Up to a shift proportional to the identity, the most general matrix $X$ consistent with the isotropy constraint and Gaussianity is a matrix from the Gaussian elliptic ensemble (see appendix~\ref{sec:iso}), i.e. a random matrix distributed with respect to the density~\cite{SCSS:1988}
\begin{equation}\label{intro:elliptic}
P^\beta_\tau(X)=\frac1Z\exp\bigg[-\frac{\beta}{2(1-\tau^2)}\tr\Big(XX^\dagger-\frac{\tau}{2}(X^2+X^{\dagger\,2})\Big)\bigg],
\end{equation}
with $Z$ denoting the normalisation constant and $\tau\in(-1,1)$ denoting an interpolation parameter between Hermitian and skew-Hermitian matrices. In general, the matrix $X$ is non-Hermitian (Ginibre for $\tau=0$) and the random evolution matrix $\Pi_{\mu=0}(t)$ is therefore a diffusion on the general linear group, $\gGl_N$. The special case is $\tau=-1$, where $X$ is skew-Hermitian and thus $\Pi_{\mu=0}(t)$ belongs to the unitary subgroup. Physically $\tau$ measures the rotational versus gradient nature of the flow, with $\tau=1$ representing a pure gradient flow.

Perhaps the simplest question we may ask about isotropic Brownian motions regards the large-time asymptotic of the strain matrix, $S(t):=\Pi_\mu^\dagger(t)\Pi_\mu(t)$. It is a well-known consequence of Oseledec's multiplicative ergodic theorem~\cite{Oseledec:1968} that the matrix $(\log S(t))/2t$ stabilises for almost all realisations as $t$ tends to infinity. The eigenvalues of this limiting matrix are the so-called Lyapunov exponents and are used as a measure of stability. A system~\eqref{intro:stratonovich} is said to describe an attractive trajectory if
\begin{equation}
\norm{\tilde y(t)}\to0 \qquad\text{for}\qquad t\to\infty,
\end{equation}
which holds true as long as the largest Lyapunov exponent is less than zero. Thus, the value of the largest Lyapunov exponent determines stability, while its fluctuations are essential for the stability-instability transition. We note that there exist a few general theorems claiming Gaussian fluctuations of the largest Lyapunov exponent, see e.g.~\cite{LePage:1982}. However, none of these are directly applicable in the limit of high dimensionality, $N\to\infty$. The heuristic understanding of this breakdown is that the theorems implicitly require a gap between the largest and second largest Lyapunov exponent, while the Lyapunov spectrum often becomes continuous when $N\to\infty$.

For isotropic Brownian motions over real vector fields, explicit formulae for the Lyapunov spectrum were obtained 30 years ago~\cite{LeJan:1985,BH:1986,Newman:1986}. However, these exponents do not capture the fluctuations in the dynamics at finite times.
A more challenging problem is to study the statistical properties of the spectrum of the strain matrix as a function of time, and thereby gain information about the finite-time Lyapunov exponents. Such results can be used to further study the double scaling limit where both the time $t$ and the dimension $N$ tend to infinity. This adds interest since non-trivial scaling regimes are expected.

The remainder of this paper is organised as follows. In section~\ref{sec:FP} we show how to pass from the matrix-valued It\^o equation \eqref{intro:ito} to the Fokker--Planck equation for the finite-time Lyapunov exponents, including the case of complex fields. In section~\ref{sec:solve} we show that the Fokker--Planck equation coincides with the one encountered in the phase-coherent transport problem, which is exactly solvable in the complex case, and provide an explicit, compact expression for the joint probability density function. Based on this expression, we also explain the connection to biorthogonal ensembles and Hamiltonian models with an external source.
The final section is devoted to the discussion in terms of May--Wigner stability analysis, and a description of further open problems and possible applications. In the appendix we recall the relation between isotropy and the Gaussian elliptic ensemble.

\section{Fokker--Planck equation for finite-time Lyapunov exponents}
\label{sec:FP}

Our goal in this section is to find the Fokker--Planck equation for the finite-time Lyapunov exponents. To do this, we need to look at the time evolution of the eigenvalues of the strain matrix $S(t)$. The easiest way to proceed is to use the product formulation~\eqref{intro:matrix-product}.

Let us denote the eigenvalues of the strain matrix by $s_1(t),\ldots,s_N(t)$. Since the strain matrix is Hermitian, it follows from ordinary perturbation theory that
\begin{equation}\label{FP:perturb-1}
\delta s_i(t):=s_i(t+\delta t)-s(t)=\delta S_{ii}+\sum_{j\neq i}\frac{\delta S_{ij}\delta S_{ji}}{s_i(t)-s_j(t)}
+\text{higher orders}
\end{equation}
where, according to~\eqref{intro:matrix-product}, we have
\begin{equation}\label{FP:perturb-2}
\delta S_{ij}=\sigma\,\delta t^{1/2}(s_jX^*_{ji}+s_iX_{ij})
+\sigma^2\delta t\Big(2\mu\,s_i\delta_{ij}+\sum_k \big(s_kX^*_{ki}X_{kj}-s_i\E[X_{ik}X_{kj}]\big)\Big)+O(\delta t^{3/2}).
\end{equation}
If we consider  evolutions over real vector fields  we have $X^*=X$, but we keep the  notation general to also account for evolutions over complex fields.

In order to find the Fokker--Planck equation for the eigenvalues of the strain matrix, 
we use the product formulation~\eqref{intro:matrix-product} and write down a recursive formula for the matrix density of the random evolution matrix $\Pi_\mu(n\delta t)$. Upon expansion in $\delta t$ and taking the limit $\delta t\to 0$ ($n\to\infty$), this recursion produces a differential equation for the matrix density. Finally, the Fokker--Planck equation is obtained by taking the expectation value of the empirical density with respect to the matrix density. This gives
\begin{equation}\label{FP:FP-general}
\frac{\p}{\p t}\rho_t(s_1,\ldots,s_N)=\lim_{\delta t\to0}
\sum_{i=1}^N\frac{\p}{\p s_i}\bigg(-\E\Big[\frac{\delta s_i}{\delta t}\Big]
+\frac12\sum_{j=1}^N\frac{\p}{\p s_j}\E\Big[\frac{\delta s_i\delta s_j}{\delta t}\Big]\bigg)
\rho_t(s_1,\ldots,s_N),
\end{equation}
where $\rho_t(s_1,\ldots,s_N)$ is the joint probability density function for the eigenvalues at time $t$. 
The next step is to note that~\eqref{FP:perturb-1} and~\eqref{FP:perturb-2} imply
\begin{equation}
\lim_{\delta t\to0}\E\Big[\frac{\delta s_i}{\delta t}\Big]=
\sum_js_j\E[X^*_{ji}X_{ji}]+\sum_{j\neq i}
\frac{s_i^2\E[X^*_{ij}X_{ij}]+s_j^2\E[X^*_{ji}X_{ji}]+2s_is_j\Re\E[X_{ij}X_{ji}]}{s_i-s_j}
\end{equation}
and
\begin{equation}
\lim_{\delta t\to0}\E\Big[\frac{\delta s_i\delta s_j}{\delta t}\Big]=
2s_is_j(\Re\E[X^*_{ii}X_{jj}]+\Re[X_{ii}X_{jj}]),
\end{equation}
which by insertion in~\eqref{FP:FP-general}, at least in principle, provides an explicit expression for the Fokker--Planck equation for the eigenvalues of our strain matrix.

With the random matrix $X$ distributed according to the density~\eqref{intro:elliptic}, we have covariances
\begin{align}
\beta=1:\qquad \E[X_{ij}X_{k\ell}]&=\delta_{ik}\delta_{j\ell}+\tau\delta_{i\ell}\delta_{jk} \qquad \text{(with $X_{ij}^*=X_{ij}$)},\\
\beta=2:\qquad \E[X^*_{ij}X_{k\ell}]&=\delta_{ik}\delta_{j\ell},\qquad
\E[X_{ij}X_{k\ell}]=\E[X^*_{ij}X^*_{k\ell}]=\tau\delta_{i\ell}\delta_{jk}.
\end{align}
Introducing these covariances into the above-given formulae, we obtain the Fokker--Planck equation
\begin{multline}\label{FP:FP-lambda}
\frac{\p}{\p t}\rho_t^\beta(s_1,\ldots,s_N)=\frac{4\kappa}{\beta}
\sum_{i=1}^N\frac\p{\p s_i}\bigg(\frac{(2-\beta+N)\beta}{2} s_i
-\beta s_i^2\sum_{j\neq i}\frac{1}{s_i-s_j}
+s_i^2\frac\p{\p s_i}\bigg)\rho_t^\beta(s_1,\ldots,s_N)\\
+2\mu\sum_{i=1}^N\frac\p{\p s_i}s_i\rho_t^\beta(s_1,\ldots,s_N).
\end{multline}
Here, the first line on the right-hand side represents a diffusive term with diffusion constant given as $\kappa=(1+\tau)\sigma^2/2$,
while the second line represents a drift term.

The final step is a change of variables from the eigenvalues of the strain matrix to the exponents $\{\lambda_k:=\frac12\log s_k\}$. After some standard manipulations, we find the Fokker--Planck equation for the exponents,
\begin{equation}\label{FP:FP-x}
\frac{\p}{\p t}\tilde\rho_t^\beta(\lambda_1,\ldots,\lambda_N)=\frac{\kappa}{\beta}
\sum_{i=1}^N\frac\p{\p \lambda_i}\bigg(\frac\p{\p\lambda_i}+
\beta\frac{\p\Omega(\lambda)}{\p\lambda_i}\bigg)\tilde\rho_t^\beta(\lambda_1,\ldots,\lambda_N)
+\mu\sum_{i=1}^N\frac\p{\p\lambda_i}\tilde\rho_t^\beta(\lambda_1,\ldots,\lambda_N)
\end{equation}
with repulsion term and initial condition given by
\begin{equation}\label{FP:Omega+boundary}
\exp[-\Omega(\lambda)]=\prod_{1\leq i<j\leq N}\sinh(\lambda_j-\lambda_i)
\qquad\text{and}\qquad
\lim_{t\to0}\tilde\rho_t^\beta(\lambda_1,\ldots,\lambda_N)=\prod_{i=1}^n\delta(\lambda_i),
\end{equation}
respectively. Here, the latter condition originates from the fact that the random evolution matrix is required to be equal to unity at $t=0$. 

% In~\eqref{FP:FP-x}, we have chosen a notation common in the study of disordered wires. 
We could alternatively write~\eqref{FP:FP-x} as
\begin{equation}\label{FP:FP-dyson}
\frac{\p\tilde\rho_t^\beta(\lambda_1,\ldots,\lambda_N)}{\p t}=
\sum_{i=1}^N\Big(\frac{\kappa}{\beta}\frac{\p^2\tilde\rho_t^\beta(\lambda_1,\ldots,\lambda_N)}{\p\lambda_i^2}
+\mu\frac{\p\tilde\rho_t^\beta(\lambda_1,\ldots,\lambda_N)}{\p\lambda_i}\Big)
-\kappa\sum_{\substack{i,j=1\\i\neq j}}^N\frac{\p}{\p\lambda_i}\bigg(\frac{\tilde\rho_t^\beta(\lambda_1,\ldots,\lambda_N)}{\tanh(\lambda_i-\lambda_j)}\bigg),
\end{equation}
which more closely resembles the notation chosen by Dyson~\cite{Dyson:1962brown}.

The Fokker--Planck equation~\eqref{FP:FP-x} is our main result in this section. In the next section, we will show that this equation is exactly solvable for $\beta=2$. However, before this a few remarks are in order.

We first note that the diffusive term in~\eqref{FP:FP-x} vanishes as $\tau\to-1$; recall that $\kappa=(1+\tau)\sigma^2/2$. This absence of diffusion  occurs since the random evolution matrix $\Pi_{\mu=0}(t)$ is unitary when $\tau=-1$. It follows that the eigenvalues of the strain matrix $S(t)=\Pi_{\mu}^\dagger(t)\Pi_{\mu}(t)$ are all identical and equal to $-\mu t$. 

As a second remark, we recall that the Lyapunov exponents $\tilde\mu_k=\lim_{t\to\infty}\lambda_k(t)/t$. In the real case ($\beta=1$) and $\mu=0$ these limiting values were originally found by Le Jan~\cite{LeJan:1985} using methods from It\^o calculus (see also~\cite{BH:1986}). An alternative approach was introduced by Newman~\cite{Newman:1986} shortly thereafter. Newman's method builds on~\cite{CN:1984} and uses isotropy to rewrite the determinant of a product of random matrices as a sum, which can then be evaluated using the law of large numbers. These  papers  focus solely on real evolutions, but the approach is easily extended to include complex matrices (see~\cite{Forrester:2013,ABK:2014,Forrester:2015,Ipsen:2015thesis} for a short discussion of such an extension in a slightly different context). In our case, the Lyapunov exponents $\tilde\mu_k$ are found to be
\begin{equation}\label{FP:lyapunov}
\tilde\mu_k=\kappa(2k-1-N)-\mu,\qquad k=1,\ldots,N.
\end{equation}
They are independent of $\beta$ and equidistantly spaced over the interval $(-\kappa N-\mu,\kappa N-\mu)$. Evidently, if we take $\sigma^2=1/N$ and thus $\kappa=(1+\tau)/2N$, we have convergence of the global spectral density to a ``square law'' on an interval of length $1+\tau$, centred at $-\mu$. This applies to the iterated limit where $t\to\infty$ followed by $N\to\infty$, for which the square law was originally pointed out by Newman~\cite{Newman:1986}. More recent results for products of random matrices lead us to believe that this law is independent of the order of the limits~\cite{Kargin:2008,ABK:2014,Ipsen:2015thesis}.

The convergence of the global spectrum allows us to establish a stability phase diagram in the large $N$ and $t$ limit. With the aforementioned scaling, the system is stable if $(1+\tau)/2<\mu$ and unstable if $(1+\tau)/2>\mu$. The Lyapunov exponents themselves can however not be used to describe the finer structure of the phase transition; this requires information about their fluctuations. The Fokker--Planck equation~\eqref{FP:FP-x} is a good starting point for a study of such fluctuations.

Heuristically, the emergence of the Lyapunov exponents~\eqref{FP:lyapunov} from the Fokker--Planck equation~\eqref{FP:FP-x} may be understood by realising that for finite $N$, the eigenvalues separate exponentially fast compared with the eigenvalue repulsion. Thus, with the ordering $\lambda_1\ll\cdots\ll\lambda_N$, in the long-time limit
\begin{equation}
\frac{\p\Omega(\lambda)}{\p\lambda_k}\approx 2k-N-1 .
\end{equation}
With this approximation the Fokker--Planck equation~\eqref{FP:FP-x} turns into $N$ uncoupled heat equations. The exponents are seen to be independently Gaussian distributed, and in the long-time limit agree with~\eqref{FP:lyapunov}.

The benefit of the Fokker--Planck equation~\eqref{FP:FP-x} is that it provides information about the statistical properties of the exponents at all times $t$. This is in contrast to Newman's method and its extensions, which only apply when $\kappa t\gg N$.

\section{Solving the Fokker--Planck equation over complex fields}
\label{sec:solve}

In this section, we show that the Fokker--Planck equation~\eqref{FP:FP-x} is exactly solvable in the complex case, i.e. for $\beta=2$. We exploit that a specific version of the Fokker--Planck equation appears for a  matrix model describing the phase-coherent transport properties of quasi-one-dimensional disordered wires with chiral symmetry~\cite{MBF:1999}. In this setting, one investigates the so-called transfer matrix $M$, which is a $2N\times 2N$ dimensional matrix that obeys the symplectic constraint $M^\dagger \sigma_1M=\sigma_1$, while chirality imposes $\sigma_3 M\sigma_3=M$ (both conditions are expressed in terms of Pauli matrices $\sigma_i$). Due to the symplectic structure, the eigenvalues $\exp(2 x_n)$ of $M^\dagger M$ occur in reciprocal pairs, $(x_n,-x_n)$. Chirality enforces a block-diagonal structure $M={\rm diag}\,(A,(A^\dagger)^{-1})$, so that the exponents $x_n$ arise from $AA^\dagger$ while the exponents $-x_{n}$ arise from $(AA^\dagger)^{-1}$. The multiplicative law of transfer matrices with Gaussian statistics \cite{MPK:1988,Beenakker:1997} then results in
the Fokker--Planck equation~\eqref{FP:FP-x} with $\tau=\mu=0$, and $\lambda_n=x_n$ identified with one of the two sets of the eigenvalues (in this context, the Fokker--Planck equation is known as the DMPK equation). 
In our case the parameters $\tau$ and $\mu$ are finite, but this can be accounted for by rescaling and shifting. 
An important difference between the  physics underlying the two models is that the transport is dominated by the transport exponent $x_n$ nearest zero, while the stability of the flow depends on the largest Lyapunov exponent $\lambda_n$.

Here, we obtain the exact solution following~\cite{MBF:1999}, and then bring the result into a compact form which more directly reveals the long time asymptotics of the Lyapunov exponents.

We first note that the drift $\mu$ only introduces an overall shift of the spectrum. This allows us to simplify notation by setting $\mu=0$ in the calculations, and then reintroduce the shift in the final result.
After this simplification, the key idea to solve~\eqref{FP:FP-x} is to parametrise the joint density as~\cite{BR:1994}
\begin{equation}\label{solve:parametrisation}
\tilde\rho_t^\beta(\lambda_1,\ldots,\lambda_N)=
\exp[-\beta(\Omega(\lambda)-\Omega(\nu))/2]\psi_t^\beta(\lambda_1,\ldots,\lambda_N\vert \nu_1,\ldots,\nu_n),
\end{equation}
where $\psi_t$ is some (wave) function, and $\nu=(\nu_1,\ldots,\nu_N)$ is a given initial condition. With this parametrisation and the notation from above, the Fokker--Planck equation turns into a  Schr\"odinger equation in imaginary time,
\begin{equation}\label{solve:schrodinger}
-\frac{\p}{\p t}\psi_t^\beta=H\psi_t^\beta,\qquad
H=-\sum_{i=1}^N\frac\kappa\beta\frac{\p^2}{\p \lambda_i^2}+V(\lambda),
\end{equation}
with the potential
\begin{equation}
V(\lambda):=\sum_{i=1}^N\frac\kappa2\Big(\frac\beta2\Big(\frac{\p\Omega(\lambda)}{\p \lambda_i}\Big)^2
-\frac{\p^2\Omega(\lambda)}{\p \lambda_i^2}\Big),
\end{equation}
which turns out to be of Calogero--Sutherland type~\cite{Calogero:1969a,Calogero:1969b,Sutherland:1971}. The evaluation of the potential is straightforward. By insertion of the definition of $\Omega(\lambda)$ from~\eqref{FP:Omega+boundary}, we find
\begin{equation}
\frac{V(\lambda)}{\kappa}=\frac{\beta-2}{4}\sum_{1\leq i<j\leq N}\coth^2(\lambda_j-\lambda_i)
+\frac{\beta}2\sum_{\substack{i,j,k=1\\ i\neq j,j\neq k,k\neq i}}^N\coth(\lambda_i-\lambda_j)\coth(\lambda_k-\lambda_j)
+\frac{\beta N(N-1)}2.
\end{equation}
Here, the second term on the right-hand side is seen to be a constant by exploiting an identity for cyclic sums. For distinct $\lambda_i,\lambda_j,\lambda_k$, we have
\begin{equation}\label{solve:cyclic}
\coth(\lambda_i-\lambda_j)\coth(\lambda_k-\lambda_j)
+\coth(\lambda_j-\lambda_k)\coth(\lambda_i-\lambda_k)
+\coth(\lambda_k-\lambda_i)\coth(\lambda_j-\lambda_i)=+1.
\end{equation}
Thus, we may write the potential as
\begin{equation}\label{solve:potential}
V(\lambda)=\frac{\kappa(\beta-2)}{4}\sum_{1\leq i<j\leq N}\coth^2(\lambda_j-\lambda_i)
+\frac{\kappa\beta(N+1)N(N-1)}{6}.
\end{equation}
Consequently the pair interaction vanishes for $\beta=2$, and the Hamiltonian in~\eqref{solve:schrodinger} becomes that of $N$ free particles. This is the feature which ensure solvability for $\beta=2$.

Now, we can return to the Schr\"odinger equation~\eqref{solve:schrodinger}. For the rest of this section we will restrict our attention to the case $\beta=2$  only, and  leave out the index for notational simplicity.

It follows from the parametrisation~\eqref{solve:parametrisation} that the wave function $\psi_t$ must be anti-symmetric in both $\lambda_1,\ldots,\lambda_N$ and $\nu_1,\ldots,\nu_N$, since the joint density $\tilde\rho_t$ is symmetric while $\Omega(\lambda)$ is anti-symmetric. Combining this with the absence of pair interactions in~\eqref{solve:schrodinger} allows us to write the wave function as a Slater determinant,
\begin{equation}\label{solve:wave}
\psi_t(\lambda_1,\ldots,\lambda_n\vert \nu_1,\ldots,\nu_N)=\frac{e^{-Ut}}{N!}\det_{1\leq i,j\leq N}[g_{j}^t(\lambda_i)],
\end{equation}
where $U=\kappa(N+1)N(N-1)/3$ is the constant contribution to the potential energy~\eqref{solve:potential}, while each $g_j^t(\lambda)$ satisfies a heat equation
\begin{equation}\label{solve:g}
\frac{\p g_j^t(\lambda)}{\p t}=\frac\kappa2\frac{\p^2g_j^t(\lambda)}{\p \lambda^2},\qquad
g_j^{t=0}(\lambda)=\delta(\lambda-\nu_j).
\end{equation}
This can be verified by inserting the wave function~\eqref{solve:wave} into the Schr\"odinger equation~\eqref{solve:schrodinger}. The solution to the heat equation~\eqref{solve:g} is a Gaussian,
\begin{equation}
g_j^t(\lambda)=\frac{1}{\sqrt{2\pi\kappa t}}\exp\Big[-\frac{(\lambda-\nu_j)^2}{2\kappa t}\Big].
\end{equation}
Now, as a consequence of~\eqref{solve:parametrisation}, we know that the joint density is
\begin{equation}\label{solve:dens-gen-bound}
\tilde\rho_t(\lambda_1,\ldots,\lambda_N)=\frac{e^{-Ut}}{N!}\det_{1\leq i,j\leq N}[g_{j}^t(\lambda_i)]
\prod_{1\leq i<j\leq N}\frac{\sinh(\lambda_j-\lambda_i)}{\sinh(\nu_j-\nu_i)},
\end{equation}
assuming a non-singular initial condition $\nu_1<\cdots<\nu_N$. We note that~\eqref{solve:dens-gen-bound} reduces to a product of Dirac delta functions in the $t\to0$ limit, as required.% At this point it is worth noting that a related result is briefly mentioned in~\cite{Biane:2009}.

We are interested in the singular initial condition $\nu_1=\cdots=\nu_N=0$, which may be obtained from~\eqref{solve:dens-gen-bound} by successive use of l'H\^opital's rule. Upon reordering of rows or columns, we find
\begin{equation}
\tilde\rho_t(\lambda_1,\ldots,\lambda_N)=e^{-Ut}
\prod_{k=1}^N\frac{e^{-\lambda_k^2/2\kappa t}}{k!\sqrt{2\pi\kappa t}}
\prod_{1\leq i<j\leq N}\Big(\frac{\lambda_j-\lambda_i}{\kappa t}\Big)\sinh(\lambda_j-\lambda_i).
\end{equation}
To further simplify the expression for the joint density, we first need to make an observation about the Lyapunov exponents from the previous section. At zero drift ($\mu=0$), we have
\begin{equation}
\sum_{j=1}^N\tilde\mu_j^2=\kappa\, U,
\end{equation}
where $\tilde\mu_j$ are the Lyapunov exponents and $U$ is the constant contribution to the potential energy as above. Now, combining this with the standard relations
\begin{equation}
\prod_{1\leq i<j\leq N}(\lambda_j-\lambda_i)=\det_{1\leq i,j\leq N}[\lambda_i^{j-1}]
\qquad\text{and}\qquad
\prod_{1\leq i<j\leq N}\sinh(\lambda_j-\lambda_i)=\det_{1\leq i,j\leq N}[\tfrac12e^{(N-2j+1)\lambda_i}],
\end{equation}
we arrive at a surprisingly simple expression for the joint density,
\begin{equation}\label{solve:jpdf}
\tilde\rho_t(\lambda_1,\ldots,\lambda_N)=\bigg(\prod_{k=1}^N\frac{1}{k!}\bigg)
\det_{1\leq i,j\leq N} \bigg[\Big(\frac{\lambda_i}{2\kappa t}\Big)^{j-1}\bigg]
\det_{1\leq i,j\leq N}\bigg[\frac{e^{-(\lambda_i-\tilde\mu_jt)^2/2\kappa t}}{\sqrt{2\pi\kappa t}}\bigg].
\end{equation}
We recall that $\kappa=(1+\tau)\sigma^2/2$, while the $\tilde\mu_j$ are given by~\eqref{FP:lyapunov}.
The joint density~\eqref{solve:jpdf} takes the form of a Vandermonde determinant times a determinant of Gaussians with means related to the Lyapunov exponents. The long-time asymptotics mentioned in the previous section thus follow straightforwardly.
Moreover, we note that even though the derivation above was performed with zero drift ($\mu=0$), the drift can at this point be reintroduced without any change to the joint density~\eqref{solve:jpdf}. The normalisation may be checked using Andreief's integration formula~\cite{Andreief:1883,deBruijn:1955}.

The joint density~\eqref{solve:jpdf} describes a special type of biorthogonal ensembles~\cite{Borodin:1998} which have been coined polynomial ensembles~\cite{KS:2014,Kuijlaars:2015}. This type of ensembles has recently gained renewed attention partly due their prominent r\^ole in the study of the singular values of random matrix products, see e.g.~\cite{AI:2015}. In the case of product ensembles, the Gaussian weights within the second determinant in~\eqref{solve:jpdf} are replaced with weights given in terms of Meijer $G$-functions. In this way, isotropic Brownian motions fit neatly into the picture of recent developments regarding products of random matrices. 

There is also a direct relation between our joint density~\eqref{solve:jpdf} and another well-known matrix ensemble, the Gaussian Unitary Ensemble (GUE) with an external source (see~\cite{BH:1996,ZinnJustin:1997,ABK:2005} with references and also~\cite{Zee:1995,JV:1996}). This ensemble is based on an $N\times N$ random Hermitian matrix, $H$, distributed with respect to the measure
\begin{equation}
P_A[dH]=\frac1Z\exp\Big[-\frac1{2\kappa t}\tr H^2+\frac1\kappa\tr HA\Big]dH,
\end{equation}
where $Z$ is a normalisation constant, $2\kappa t$ is the variance, and $A$ is an $N\times N$ Hermitian external source matrix which without loss of generality may be taken to be diagonal, $A:=\diag(a_1,\ldots,a_N)$. The joint density for the eigenvalues of $H$ is obtained by integrating out irrelevant degrees of freedom using the Harish-Chandra--Itzykson--Zuber integral~\cite{HC:1957,IZ:1980}. One then finds
\begin{equation}\label{relation:ext-jpdf}
\hat\rho_t(\lambda_1,\ldots,\lambda_N)=\frac{1}{\tilde Z}
\det_{1\leq i,j\leq N} \bigg[\Big(\frac{\lambda_i}{2\kappa t}\Big)^{j-1}\bigg]
\det_{1\leq i,j\leq N}\bigg[\frac{e^{-(\lambda_i-a_j t)^2/2\kappa t}}{\sqrt{2\pi\kappa t}}\bigg]
\end{equation}
with $\lambda_1,\ldots,\lambda_N$ denoting the eigenvalues of $H$ and $\tilde Z$ representing the corresponding normalisation constant. Here, it is assumed that $a_i\neq a_j$ for $i\neq j$.

The similarity between~\eqref{relation:ext-jpdf} and~\eqref{solve:jpdf} is immediately recognised. Thus the Lyapunov exponents in~\eqref{solve:jpdf} may be reinterpreted as equidistantly spaced eigenvalues of an external source matrix, as considered e.g. in~\cite{CW:2014}. While relations between ensembles with an external source and non-intersecting Brownian motions are not new (see~\cite{ABK:2005} and references within), the isotropic Brownian motions studied in this paper provide an example where the external source arises naturally; rather than from an imposed boundary condition.

\section{Conclusions and open problems}

In this paper we have studied a family of stochastic processes with isotropic matrix-valued multiplicative noise. We have shown that it is possible to formulate a Fokker--Planck equation for the finite-time Lyapunov exponents, and that this Fokker--Planck equation is exactly solvable for evolutions over complex fields, where they  give rise to a biorthogonal ensemble. We motivated this stochastic process by its relation to a May--Wigner-like stability analysis for trajectories in an $N$-dimensional space, characterized by a mean relaxation rate $\mu$, a noise strength $\sigma$, and a parameter $\tau\in(-1,1)$ which characterizes the rotational nature of the flow ($\tau=1$ represents a pure gradient flow). For a noise variance $\sigma^2=1/N$, the infinite-time Lyapunov spectrum has compact support, and converges to a uniform distribution on an interval with length $1+\tau$ centred at $-\mu$. This allows us to establish a phase diagram for the system in the large-$N$ limit, according to which trajectories are  stable if $(1+\tau)/2<\mu$ and unstable if $(1+\tau)/2>\mu$.

The heuristic argument at the end of section~\ref{sec:FP} suggests that the fluctuations of the largest Lyapunov exponent become Gaussian in the limit $\kappa t\gg N$, with more rigorous formulations of this  statement following from~\cite{LeJan:1985,CN:1984}. It is more challenging to investigate the fluctuations away from this limit, where they are expected to be non-Gaussian. Here, the Fokker--Planck equation established in this paper provides a good starting point. In particular, we note that $\kappa t\ll N$ results in a limit where adjacent stability exponents are close compared to their correlations length. With this in mind, we may substitute $\tanh(\lambda_j-\lambda_i)\approx(\lambda_j-\lambda_i)$ in~\eqref{FP:FP-dyson},  which transforms the Fokker--Planck equation into an ordinary Dyson diffusion. This suggests that the largest Lyapunov exponent follows the so-called Tracy--Widom law. Similar approximations may be made starting with the joint density function from section~\ref{sec:solve}. We note that this conjectural transition from a classical random matrix law (Tracy--Widom distribution) to a Gaussian law is not completely unfamiliar; in~\cite{JQ:2014} it was shown that the largest eigenvalue (in terms of absolute value) for a product of complex Ginibre matrices undergoes a transition from a Gumbel distribution to a log-normal distribution. 

Future work may be directed towards establishing this transition, and extending it to a detailed description of the 
stability-instability phase transition, including the critical exponents. 
From a more mathematical perspective, there are many other intriguing questions worth pursuing beyond the fluctuations of the largest Lyapunov exponent. This includes, but is not limited to, a study of the global spectrum as a function of time, and the local correlations in the bulk as well as near the edge at fixed time $t$.

\paragraph{Acknowledgement:} We like to thank G. Akemann, P. J. Forrester, and M. Kieburg for useful discussions. JRI acknowledge financial support by ARC Centre of Excellence for Mathematical and Statistical Frontiers.

\appendix

\section{Isotropic measures and elliptic ensembles}
\label{sec:iso}

In this appendix we briefly recall the construction of isotropic Gaussian matrix measures and their relation to Gaussian elliptic ensembles. We will consider the real ($\beta=1$) and the complex ($\beta=2$) case separately.

First, let us look at real matrices; our description closely follows that of~\cite{LeJan:1985}. We are considering the family of Gaussian probability measures on the space of $N\times N$ real matrices,  where each entry is (centred) Gaussian distributed. Any such measure is uniquely determined by its covariance tensor, $\E[X_{ij}X_{k\ell}]$. Imposing isotropy implies that the measure must be invariant under all orthogonal similarity transformations, $X\mapsto U^TXU$. On the level of the covariance tensor, this statement becomes
\begin{equation}
\E[X_{i'j'}X_{k'\ell'}]=\sum_{i,j,k,\ell=1}^N U_{i'i}U_{j'j}U_{k'k}U_{\ell'\ell}\E[X_{ij}X_{k\ell}]
\end{equation}
for all $U\in\gO(N)$. Consequently, any isotropic measure has a covariance tensor given by
\begin{equation}
\E[X_{ij}X_{k\ell}]=a\,\delta_{ij}\delta_{k\ell}+b\,\delta_{ik}\delta_{j\ell}+c\,\delta_{i\ell}\delta_{jk}
\end{equation}
with $a,b,c$ denoting constants unaffected by orthogonal similarity transformations. To understand the interpretation of these constants, we note that
\begin{equation}
\E[(X_{ii})^2]=a+b+c\geq0\qquad\text{and}\qquad
\frac12\E[(X_{ij}\pm X_{ji})^2]=b\pm c\geq0\qquad(i\neq j).
\end{equation}
Thus, we may write the random matrix $X$ as a sum
\begin{equation}
X=\sqrt{\frac{b+c}{2}}H+\sqrt{\frac{b-c}{2}}A+\sqrt{a}\,\xi\one,
\end{equation}
where $H$ is a symmetric matrix with standard Gaussian entries (i.e. GOE), $A$ is a skew-symmetric matrix with standard Gaussian entries (i.e. skew-GOE), and $\xi$ is a standard Gaussian random variable. Setting $a=0$, $b=1$, and $c=\tau\in(-1,1)$ result in the elliptic density~\eqref{intro:elliptic} with $\beta=1$.

We can now turn to complex matrices ($\beta=2$). Here, we need to take two covariance tensors into account when determining the most general Gaussian measure, $\E[X_{ij}^*X_{k\ell}]$ and $\E[X_{ij}X_{k\ell}](=\E[X_{ij}^*X_{k\ell}^*])$. Isotropy says that the measure should be invariant under unitary similarity transformations, $X\mapsto U^\dagger XU$, which means that
\begin{align}
\E[X_{i'j'}X_{k'\ell'}]&=\sum_{i,j,k,\ell=1}^N U_{i'i}^*U_{j'j}U_{k'k}^*U_{\ell'\ell}\E[X_{ij}X_{k\ell}],\\
\E[X_{i'j'}^*X_{k'\ell'}]&=\sum_{i,j,k,\ell=1}^N U_{i'i}U_{j'j}^*U_{k'k}^*U_{\ell'\ell}\E[X_{ij}^*X_{k\ell}],
\end{align}
for all $U\in\gU(N)$. Thus
\begin{equation}
\E[X_{ij}X_{k\ell}]=a\,\delta_{ij}\delta_{k\ell}+b\,\delta_{i\ell}\delta_{jk}
\qquad\text{and}\qquad
\E[X_{ij}^*X_{k\ell}]=c\,\delta_{ij}\delta_{k\ell}+d\,\delta_{ik}\delta_{j\ell}
,
\end{equation}
where $a,b,c,d$ are constants. Similar to the real case, we will look at correlations to obtain an interpretation of these constants. We then find
\begin{equation}
\frac12\E[\abs{X_{ii}\pm X_{ii}^*}^2]=c+d\pm (a+b)\geq0\qquad\text{and}\qquad
\frac12\E[\abs{X_{ij}\pm X_{ji}^*}^2]=d\pm b\geq0\qquad(i\neq j),
\end{equation}
and can therefore write the random matrix $X$ as a sum
\begin{equation}
X=\sqrt{\frac{d+b}{2}}H+\sqrt{\frac{d-b}{2}}A+(\sqrt{c}\,\xi+i\sqrt a\,\eta)\one,
\end{equation}
where $H$ is a Hermitian matrix with standard complex (real on the diagonal) Gaussian entries (i.e. GUE), $A$ is a skew-Hermitian matrix with standard complex (imaginary on the diagonal) Gaussian entries (i.e. skew-GUE), while $\xi$ and $\eta$ are standard real Gaussian random variables. Analogously to before we set $a=0$, $b=\tau\in(-1,1)$, $c=0$, and $d=1$, which results in a random matrix with density~\eqref{intro:elliptic}.

%% Bibliography -------------------------------------------------

\raggedright

\bibliographystyle{unsrt}
\bibliography{ref}

\end{document}